\documentclass{moriond}

\def\Journal#1#2#3#4{{#1} {\bf #2}, #3 (#4)}

\def\NIMA{{\em Nucl. Instrum. Methods} A}

\def\PLB{{\em Phys. Lett.}  B}
\def\PRL{\em Phys. Rev. Lett.}
\def\PRD{{\em Phys. Rev.} D}

\def\be{\begin{equation}}
\def\ee{\end{equation}}
\def\bea{\begin{eqnarray}}
\def\eea{\end{eqnarray}}

\newcommand{\Photo}{\includegraphics[height=35mm]{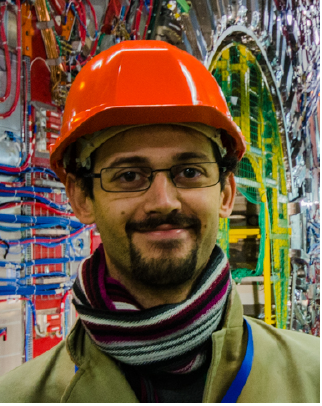}}

\usepackage{ifthen}
\usepackage{xspace}
\newboolean{uprightparticles}
\setboolean{uprightparticles}{false}

\ifthenelse{\boolean{uprightparticles}}%
{

 \def\PDelta      {\ensuremath{\Delta}\xspace}                 
 \def\PXi      {\ensuremath{\Xi}\xspace}                 
 \def\PLambda      {\ensuremath{\Lambda}\xspace}                 
 \def\PSigma      {\ensuremath{\Sigma}\xspace}                 
 \def\POmega      {\ensuremath{\Omega}\xspace}                 
 \def\PUpsilon      {\ensuremath{\Upsilon}\xspace}                 
 
 \def\PB      {\ensuremath{\mathrm{B}}\xspace}                 
                  
 \def\PD      {\ensuremath{\mathrm{D}}\xspace}

 \def\PK      {\ensuremath{\mathrm{K}}\xspace}

 \def\Pi      {\ensuremath{\mathrm{i}}\xspace}

}
{

 \mathchardef\PDelta="7101
 \mathchardef\PXi="7104
 \mathchardef\PLambda="7103
 \mathchardef\PSigma="7106
 \mathchardef\POmega="710A
 \mathchardef\PUpsilon="7107
                  
 \def\PB      {\ensuremath{B}\xspace}                 
                  
 \def\PD      {\ensuremath{D}\xspace}

 \def\PK      {\ensuremath{K}\xspace}

 \def\Pi      {\ensuremath{i}\xspace}

}

\def\kaon  {\ensuremath{\PK}\xspace}
  \def\Kbar  {\kern 0.2em\overline{\kern -0.2em \PK}{}\xspace}

\def\KS    {\ensuremath{\kaon^0_{\rm\scriptscriptstyle S}}\xspace}

  \def\Dbar    {\kern 0.2em\overline{\kern -0.2em \PD}{}\xspace}

\def\Bbar    {\ensuremath{\kern 0.18em\overline{\kern -0.18em \PB}{}}\xspace}

  \def\Y#1S{\ensuremath{\PUpsilon{(#1S)}}\xspace}

\def\Lbar {\ensuremath{\kern 0.1em\overline{\kern -0.1em\PLambda}}\xspace}

\def\to                 {\ensuremath{\rightarrow}\xspace}

\def\AT#1     {\ensuremath{A_{\mathrm{T}}^{#1}}\xspace}           

\def\C#1      {\ensuremath{\mathcal{C}_{#1}}\xspace}                       
\def\Cp#1     {\ensuremath{\mathcal{C}_{#1}^{'}}\xspace}                    
\def\Ceff#1   {\ensuremath{\mathcal{C}_{#1}^{\mathrm{(eff)}}}\xspace}        
\def\Cpeff#1  {\ensuremath{\mathcal{C}_{#1}^{'\mathrm{(eff)}}}\xspace}       
\def\Ope#1    {\ensuremath{\mathcal{O}_{#1}}\xspace}                       
\def\Opep#1   {\ensuremath{\mathcal{O}_{#1}^{'}}\xspace}                    

\newcommand{\tev}{\ifthenelse{\boolean{inbibliography}}{\ensuremath{~T\kern -0.05em eV}\xspace}{\ensuremath{\mathrm{\,Te\kern -0.1em V}}\xspace}}
\newcommand{\gev}{\ensuremath{\mathrm{\,Ge\kern -0.1em V}}\xspace}
\newcommand{\mev}{\ensuremath{\mathrm{\,Me\kern -0.1em V}}\xspace}
\newcommand{\kev}{\ensuremath{\mathrm{\,ke\kern -0.1em V}}\xspace}
\newcommand{\ev}{\ensuremath{\mathrm{\,e\kern -0.1em V}}\xspace}
\newcommand{\gevc}{\ensuremath{{\mathrm{\,Ge\kern -0.1em V\!/}c}}\xspace}
\newcommand{\mevc}{\ensuremath{{\mathrm{\,Me\kern -0.1em V\!/}c}}\xspace}
\newcommand{\gevcc}{\ensuremath{{\mathrm{\,Ge\kern -0.1em V\!/}c^2}}\xspace}
\newcommand{\gevgevcccc}{\ensuremath{{\mathrm{\,Ge\kern -0.1em V^2\!/}c^4}}\xspace}
\newcommand{\gevgevcc}{\ensuremath{{\mathrm{\,Ge\kern -0.1em V^2\!/}c^2}}\xspace}
\newcommand{\mevcc}{\ensuremath{{\mathrm{\,Me\kern -0.1em V\!/}c^2}}\xspace}

\def\gsim{{~\raise.15em\hbox{$>$}\kern-.85em
          \lower.35em\hbox{$\sim$}~}\xspace}
\def\lsim{{~\raise.15em\hbox{$<$}\kern-.85em
          \lower.35em\hbox{$\sim$}~}\xspace}

\def\tell1  {TELL1\xspace}
\def\ukl1   {UKL1\xspace}

\begin{document}

\vspace*{4cm}
\title{An analysis of $\Lambda_b^0 \rightarrow \Lambda\mu^+\mu^-$ decays at the LHCb experiment}

\author{ L. Pescatore$^*$, \\
		  on behalf of the LHCb collaboration }

\address{$^*$University of Birmingham, United Kingdom}


\maketitle
\abstract{
  The branching fraction of the rare decay $\Lambda_b^0 \rightarrow \Lambda \mu^+ \mu^-$
  is measured as a function of $q^2$, the square of the dimuon invariant mass.  The analysis is performed using
  proton-proton collision data, corresponding to an integrated luminosity of 3.0 fb$^{-1}$, collected by the LHCb experiment. 
  Evidence of signal is found for the first time in the $q^2$ region below the square of the J/$\psi$ mass. 
  In the $q^2$ intervals where the signal is observed, angular distributions are studied
  and two forward-backward asymmetries, in the dimuon and hadronic systems, are measured for the first time.
}

\section{Rare decays and $\Lambda_b^0$}

The $\Lambda_b^0 \rightarrow \Lambda \mu^+\mu^-$ decay is a rare ($b \rightarrow s$) flavour changing transition,
which is forbidden at tree level in the Standard Model (SM) but can happen through loop level electroweak processes~\cite{gutsche}.
Since the branching ratio of this type of decays is small, typically $\sim 10^{-6}$ or less,
they are very sensitive to contributions from physics beyond the SM.
The study of $\Lambda^0_b$ decays is of particular interest for severals reasons. First of all, the $\Lambda^0_b$ baryon
has non-zero spin, which allows to extract additional information about the helicity structure of the underlying theory.
Secondly, the $\Lambda^0_b$ baryon can be considered as a heavy quark plus a light di-quark system,
and therefore the hadronic physics differs significantly from similar $B$ meson decays.
Finally, a further motivation specific to the $\Lambda_b^0 \rightarrow\Lambda\mu^+\mu^-$ decay is that
the $\Lambda$ baryon decays weakly and its polarisation is preserved which
gives access to complementary information to that available from meson decays~\cite{vandyk}.
In this work the differential branching ratio and angular observables are measured
as a function of the square of the dimuon mass, $q^2$, since theoretically different treatments
of form factors are used depending on the considered $q^2$ region~\cite{vandyk}.
Previous observations of the decay~\cite{lhcbLmumu}~\cite{cdfLmumu} $\Lambda_b^0 \rightarrow \Lambda \mu^+\mu^-$
found evidence for signal only in $q^2$ intervals above the square of the mass of the $J/\psi$ resonance.
These proceedings are based on an analysis~\cite{lastLmumu} using $pp$ collision data,
corresponding to an integrated luminosity of 3.0 fb$^{-1}$, collected during 2011 and 2012 by the
LHCb detector~\cite{lhcb_detector} at centre-of-mass energies of 7 and 8 TeV, respectively.

\section{Differential branching fraction}

In order to measure their differential branching fraction,
candidate $\Lambda_b^0 \rightarrow \Lambda \mu^+ \mu^-$ decays are reconstructed
from a dimuon and a $\Lambda$ candidate, where the $\Lambda$ baryon is reconstructed through
its $p\pi^-$ decay mode. The rates are normalised with respect to the tree level $b\rightarrow c\bar{c}s$
decay $\Lambda_b^0 \rightarrow J/\psi \Lambda$, where the $J/\psi$ meson decays into a $\mu^+\mu^-$ pair,
as this mode has the same final state particles as the signal channel.

The selection is based on a neural network classifier~\cite{nn}.
The signal sample used to train the neural network consists of simulated $\Lambda_b^0 \rightarrow \Lambda \mu^+ \mu^-$ events,
while the background is taken from data in the upper sideband of the $\Lambda_b^0$ candidate mass spectrum.
The variable that provides the greatest discrimination is the $\chi^2$ from a kinematic fit
of the complete decay chain in which the proton and pion are constrained such that the $p\pi^{-}$
invariant mass corresponds to the known $\Lambda$ baryon mass, and the $\Lambda$ and dimuon systems
are constrained to originate from their respective vertices.
Other variables that contribute significantly are: the transverse momentum of the $\Lambda$ candidate;
the particle identification information for the muons,
a likelihood variable built using information from LHCb's muon system and ring imaging
Cherenkov detectors~\cite{lhcb_detector}; the separation of the muons,
the pion and the $\Lambda$ candidate from the $pp$ interaction vertex;
and the distance between the $\Lambda_b^0$ and $\Lambda$ decay vertices.

Since the $\Lambda$ baryon is long-lived, its decay has a topology with a displaced secondary vertex.
This leads to little background from other decays: the only relevant contribution
comes $B^0$ decays into $\KS$ and muons, where $\KS \rightarrow \pi^+\pi^-$ and one pion is
misidentified as a proton. This background, especially significant for the normalisation
channel, is~modelled~in~the~fit.

\begin{figure}[hb!]
\centering
\includegraphics[width=.48\textwidth]{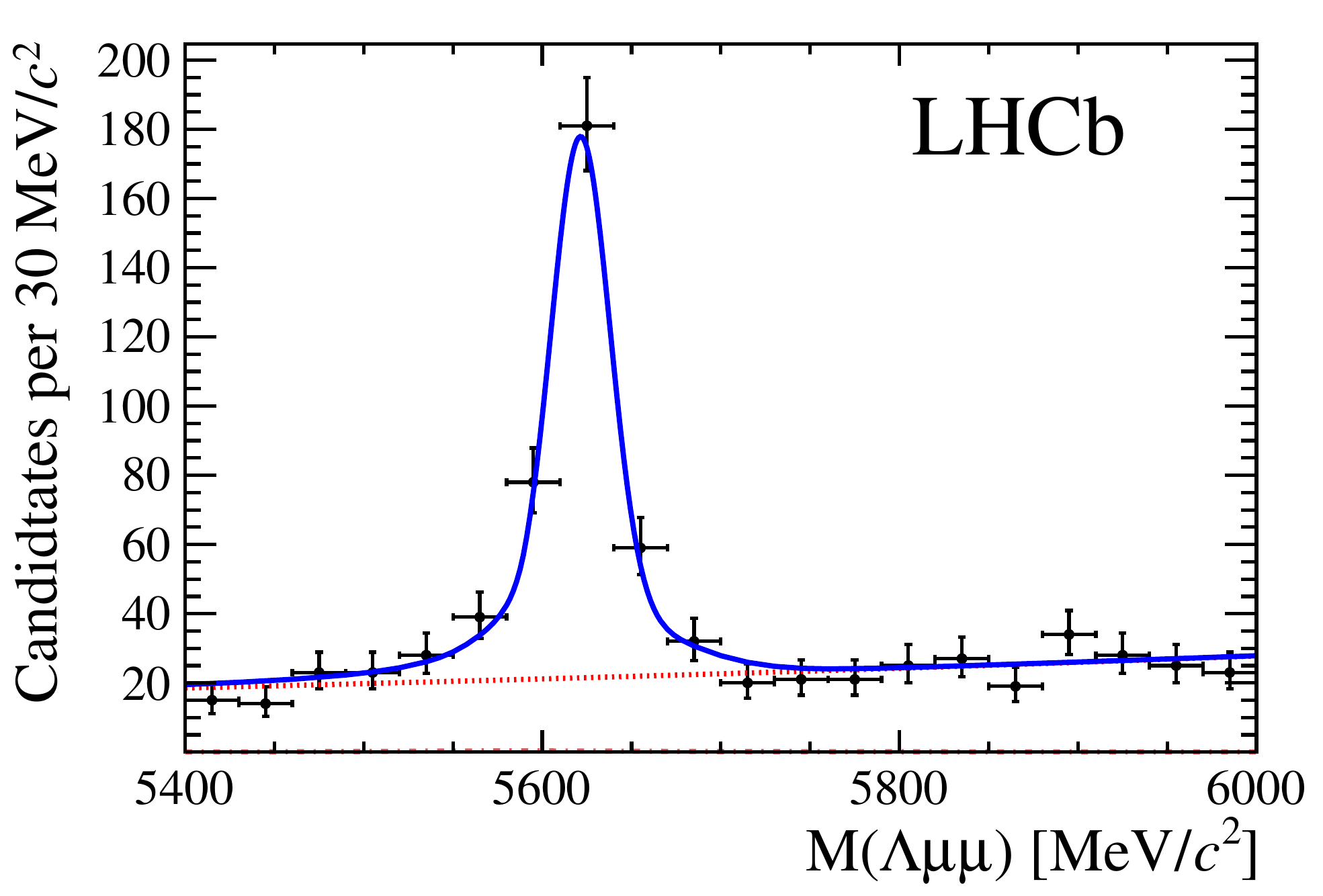}
\includegraphics[width=.48\textwidth]{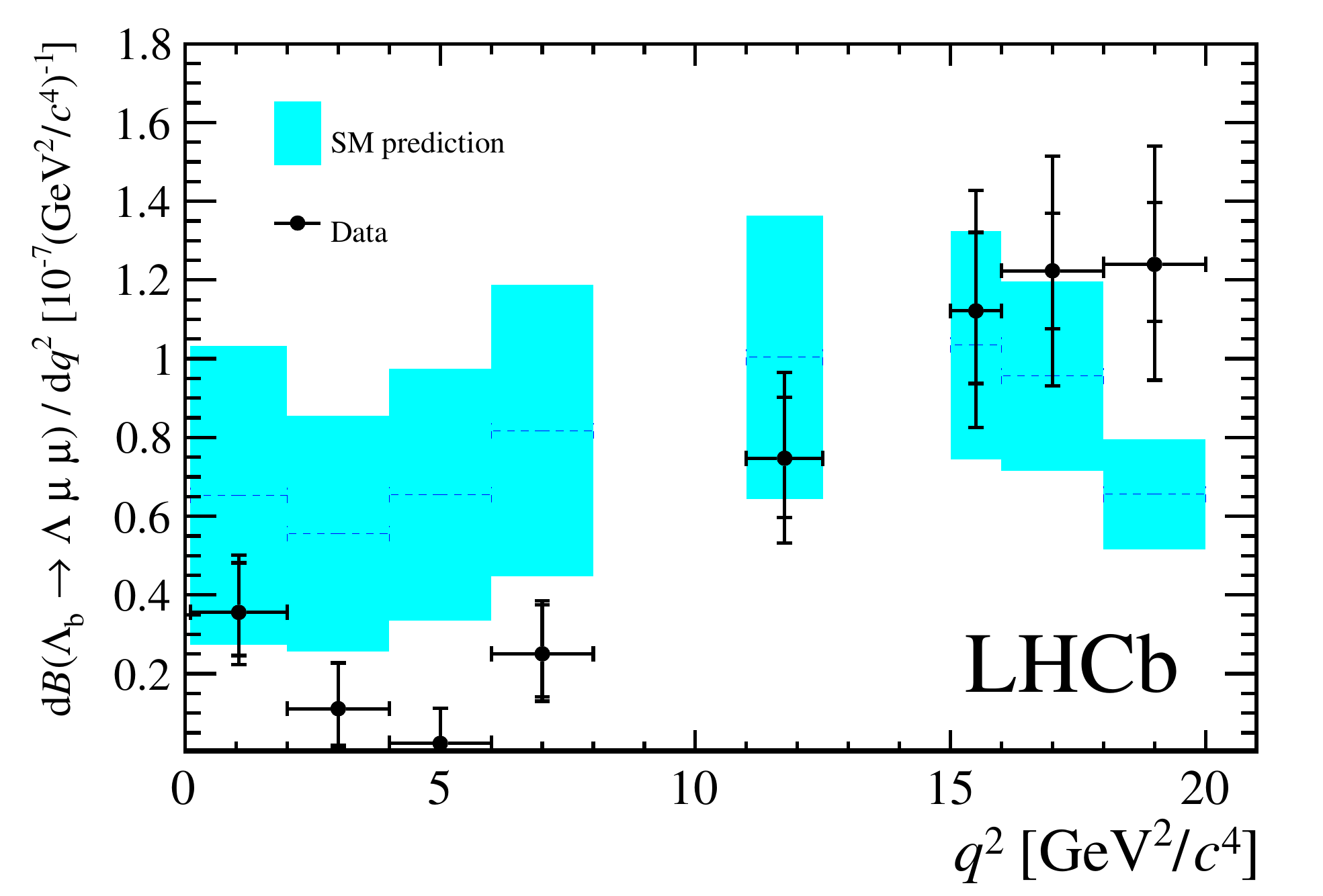}
\caption{(left) Invariant mass of the $\Lambda\mu^+\mu^-$ system in the
$15 < q^2 < 20$ GeV$^2/c^4$ interval with the fit function overlaid.
(right) Measured differential branching fraction of the rare 
$\Lambda_b^0 \rightarrow \Lambda \mu^+ \mu^-$ decay as a function of $q^2$
with the SM prediction~\protect\cite{prediction} superimposed.
The inner error bar on data points represents the total uncertainty on the relative
branching fraction (statistical and systematic); the outer error bar also includes
the uncertainties from the branching fraction of the normalisation mode.}
\label{fig:BR}
\end{figure}

In Fig.~\ref{fig:BR} the invariant mass distribution of the $\Lambda\mu^+\mu^-$ system
is reported in the $q^2$ interval $15 < q^2 < 20$ GeV$^2/c^4$ together with the fit function
used to extract the yield. The signal is modelled using the sum of two Crystal Ball functions
and the combinatorial background with an exponential.
The background from $\KS$ decays is very small in the rare decay sample and is not visible
on Fig.~\ref{fig:BR}.

The absolute branching fraction of the $\Lambda_b^0 \rightarrow \Lambda \mu^+ \mu^-$ decay
is obtained by multiplying the relative branching fraction by the absolute branching fraction of the
normalisation channel~\cite{pdg}, $\mathcal{B}(\Lambda_b^0 \rightarrow J/\psi \Lambda) = (6.3 \pm 1.3) \times 10^{-4}$.
Measured values are given in Fig.~\ref{fig:BR} as a function of $q^2$ together with SM predictions~\cite{prediction}.
The uncertainty on these values is dominated by the precision of the branching fraction
for the normalisation channel, while the uncertainty on the relative branching fraction is dominated
by the size of the data sample available.

\section{Angular analysis}

\begin{figure}[t]
\centering
\includegraphics[width=.48\textwidth]{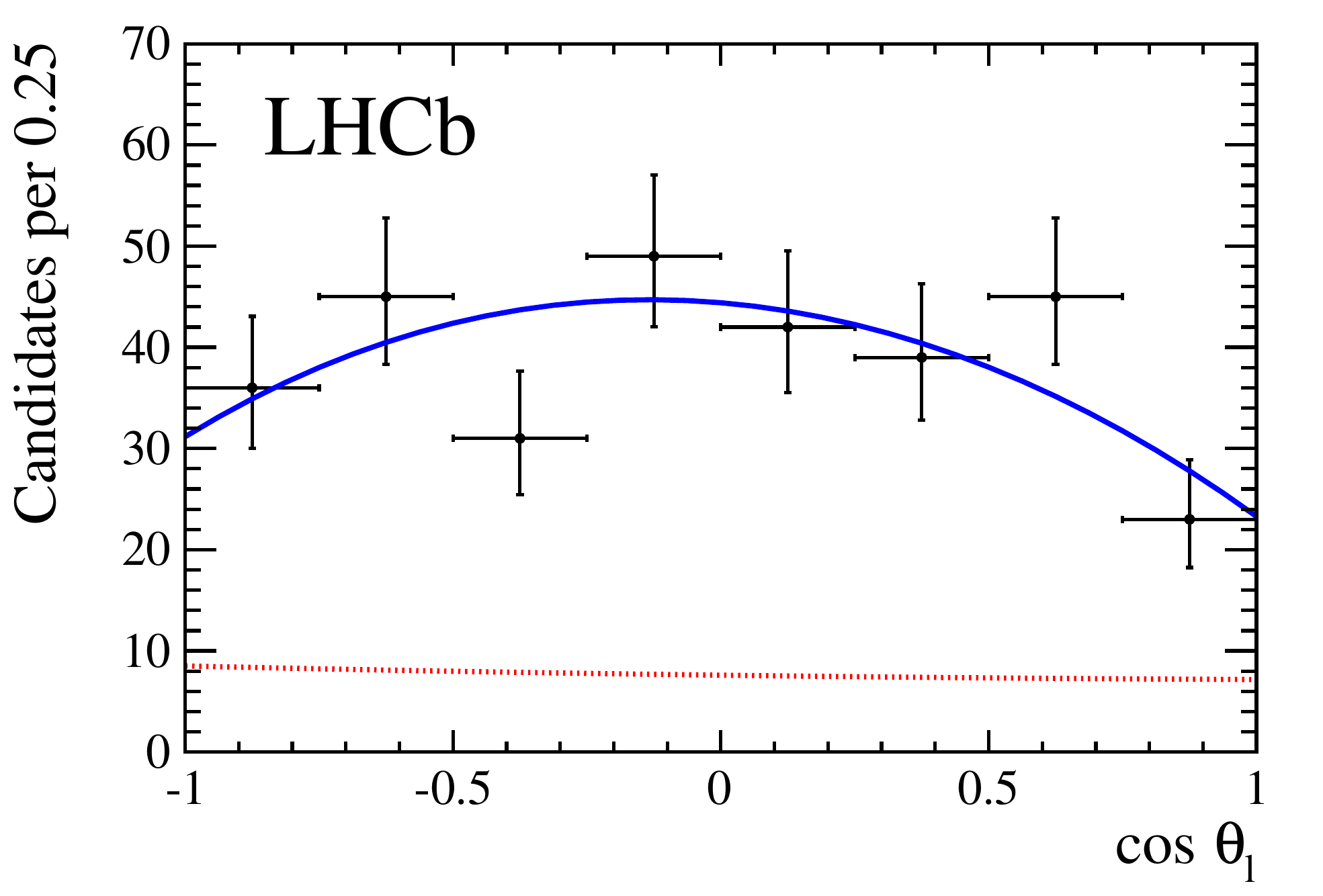}
\includegraphics[width=.48\textwidth]{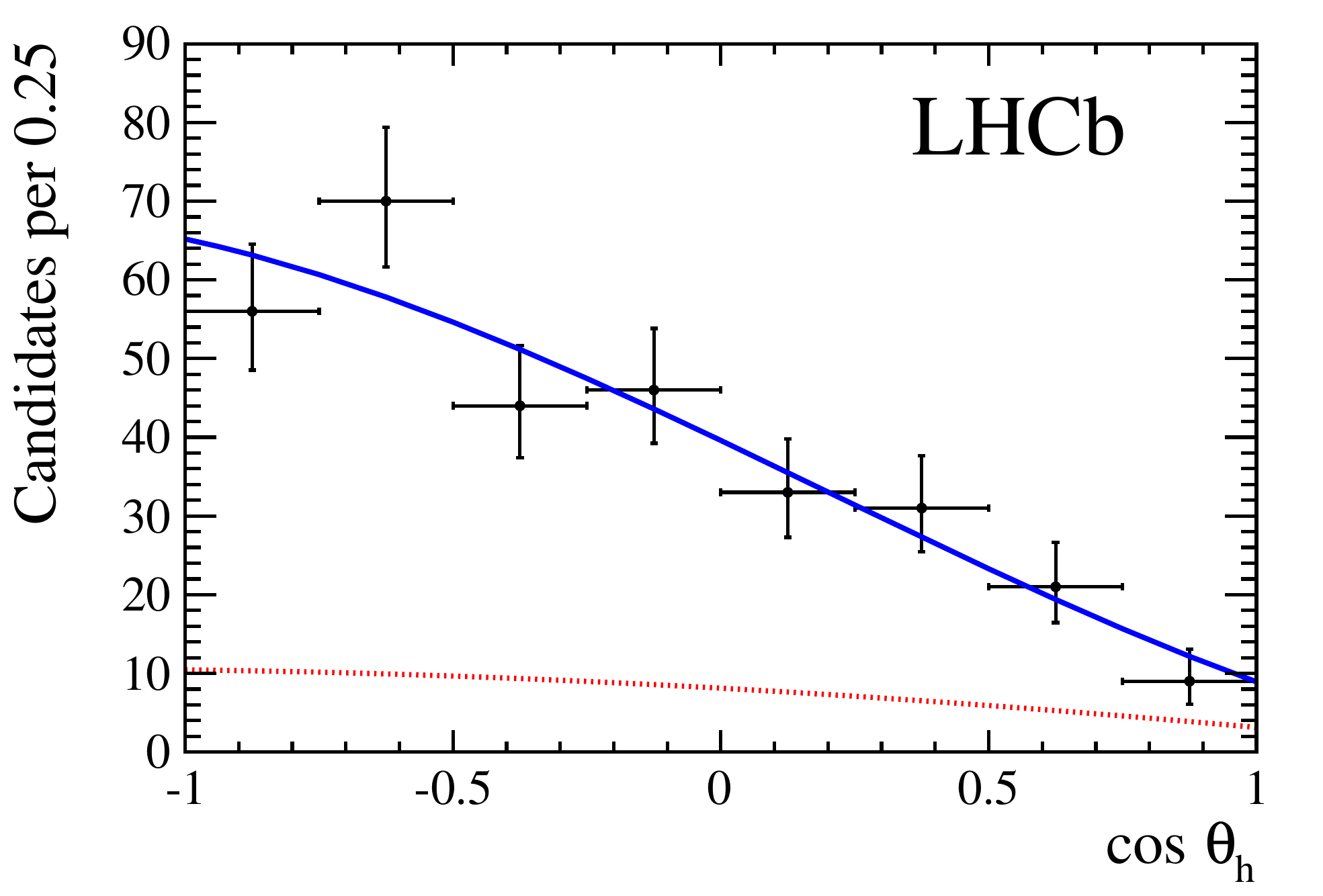}
\caption{Distributions of the (left) $\cos \theta_\ell$ and (right) $\cos \theta_h$ angles in data
in the $15 < q^2 < 20$ GeV$^2/c^4$ interval
with overlaid the total fit function, solid (blue) line, and the background component, dashed (red) line.}
\label{fig:ang_fit}
\end{figure}

\begin{figure}[bh!]
\centering
\includegraphics[width=.48\textwidth]{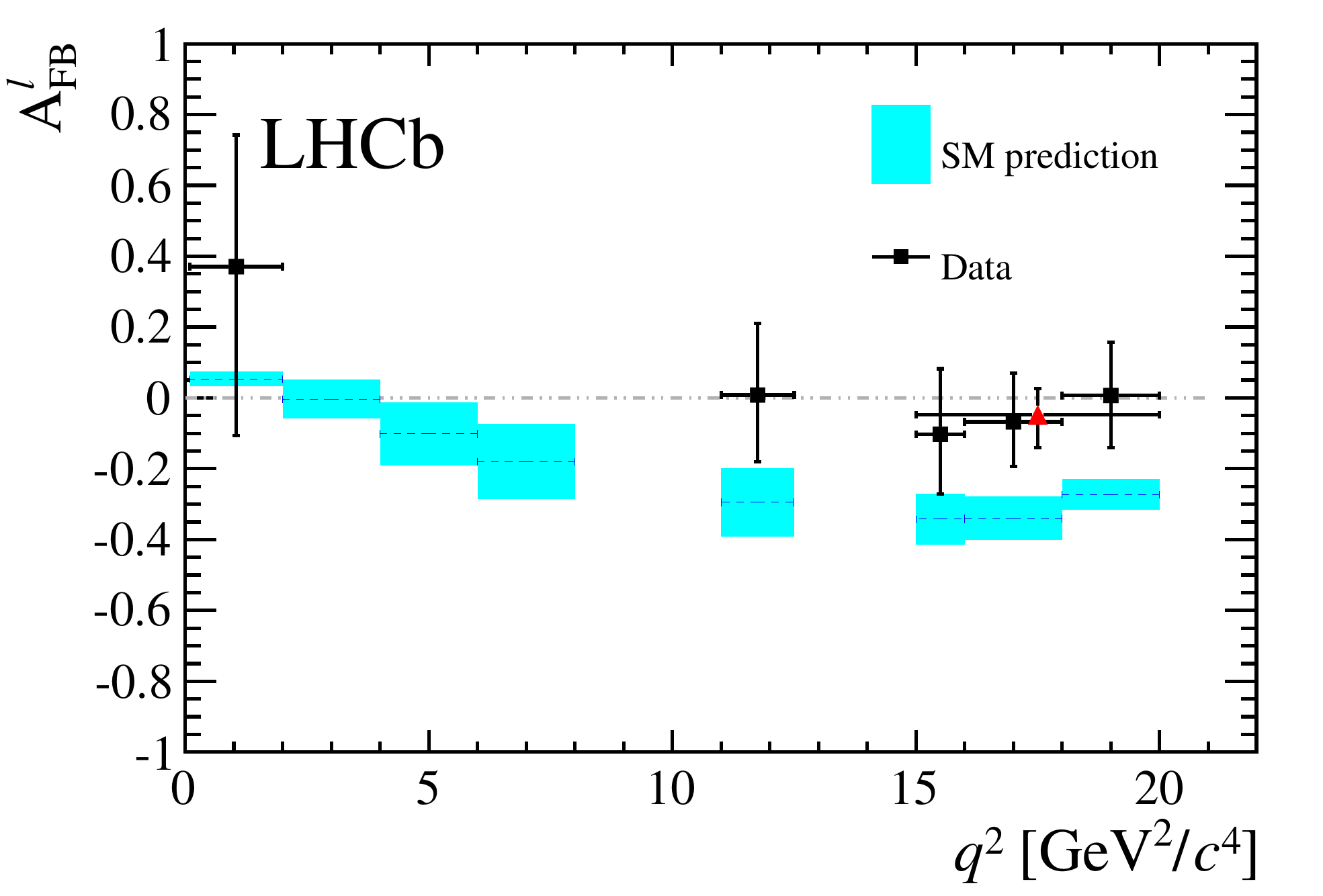}
\includegraphics[width=.48\textwidth]{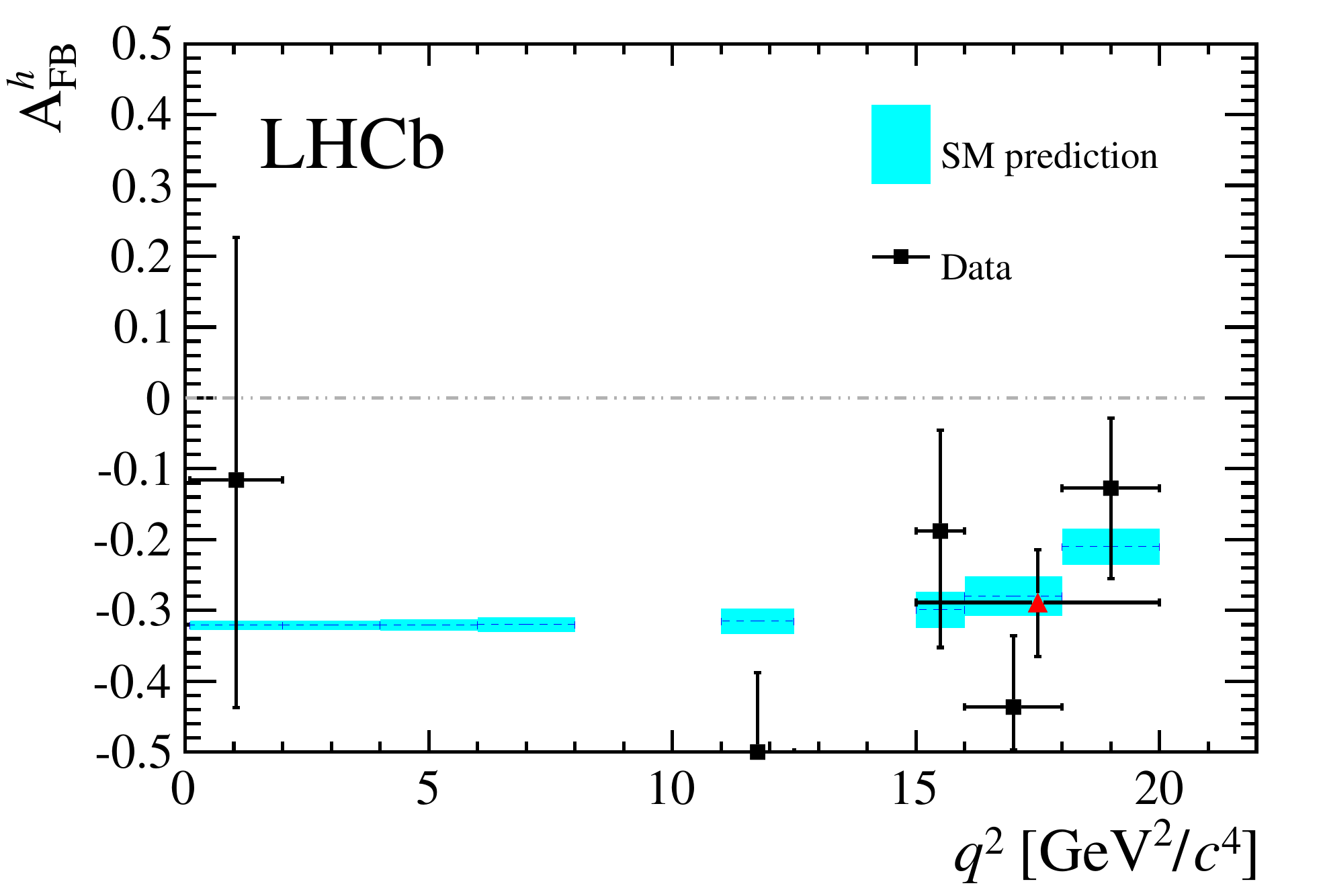}
\caption{Forward-backward asymmetries as a function of $q^2$ 
in the dimuon system (left) and $p\pi^{-}$ system (right)
with the SM prediction~\protect\cite{prediction} overlaid.}
\label{fig:ang}
\end{figure}

The $\Lambda_b^0 \rightarrow \Lambda \mu^+ \mu^-$  decay has a non-trivial angular structure which,
in the case of unpolarised $\Lambda_b^0$ production,
is described by the helicity angles of the muon ($\theta_\ell$) and proton ($\theta_h$), the angle between the planes
defined by the $\Lambda$ decay products and the two muons, and the square of the dimuon invariant mass, $q^2$.
The angle $\theta_\ell$ is defined as the angle between the positive (negative) muon and the dimuon system directions
and $\theta_h$ as the angle between the proton and the $\Lambda$ baryon directions, both in the $\Lambda_b^0$ ($\bar{\Lambda}_b^0$) rest frame.

In this work two forward-backward asymmetries, in the dimuon ($A^\ell_{\rm FB}$) and $p\pi^-$ ($A^h_{\rm FB}$) systems, are measured.
The observables are determined from one-dimensional angular distributions as a function of $\cos \theta_\ell$
and $\cos \theta_h$. The differential rate as a function of $\cos \theta_\ell$ is described by the function

\begin{equation}
\frac{\mbox{d}^2\Gamma(\Lambda_{b}\rightarrow \Lambda \,\ell^{+}\ell^{-})}{\mbox{d} q^2 \,\mbox{d}\!\cos\theta_\ell} =
\frac{\mbox{d}\Gamma}{\mbox{d} q^2}
 \left[ \frac{3}{8}\left(1+\cos^2\theta_\ell\right)(1-f_{\rm L})+A_{\rm FB}^\ell\cos\theta_\ell +
   \frac{3}{4}f_{\rm L} \sin^2\theta_\ell \right],
\label{eq:afbLTh}
\end{equation}
\noindent
where $f_{\rm L}$ is the fraction of longitudinally polarised dimuons. The rate as a function of $\cos \theta_h$ has the form
\begin{equation}
\frac{\mbox{d}^2\Gamma(\Lambda_{b}\to \Lambda(\rightarrow p \pi^{-})\ell^{+}\ell^{-})}
     {\mbox{d} q^2\,\mbox{d}\!\cos\theta_h}
={\mathcal{B}}(\Lambda \to p\pi^{-})
 \frac{\mbox{d}\Gamma(\Lambda_b \to \Lambda\, \ell^{+}\ell^{-})}{\mbox{d} q^2}\frac{1}{2}
\Big(1+2A_{\rm FB}^h\cos\theta_h\Big) \,.
\label{eq:afbBTh}
\end{equation}
\noindent
These expressions assume that $\Lambda_b^0$ baryons are produced unpolarised, which is in agreement with the
$\Lambda_b^0$ production polarisation recently measured at LHCb~\cite{lhcb_polarization}.

The observables are measured using unbinned maximum likelihood fits.
The signal PDF consists of a theoretical shape, given by Eqs.~\ref{eq:afbLTh} and~\ref{eq:afbBTh},
multiplied by a function modelling the angular efficiency.
Selection requirements on the minimum momentum of the muons generate distortions in the $\cos \theta_\ell$ distribution
by removing candidates with extreme values of $\cos \theta_\ell$. Similarly, impact parameter requirements
remove events with large $\left|\cos \theta_h\right|$, since very forward hadrons tend to have smaller impact parameter values.
The angular efficiency is parametrised using a second-order polynomial and determined by fitting simulated events.

To limit systematic uncertainties related to the background parametrisation, a narrow interval around the mass peak, dominated
by the signal, is used in the angular analysis and a polynomial component is added to the fit to account for the residual background.

Figure \ref{fig:ang_fit} shows the $\cos \theta_\ell$ and $\cos \theta_h$ distributions
in the $15 < q^2 < 20$ GeV$^2/c^4$ interval with the fit functions overlaid.
Measured values of the leptonic and hadronic forward-backward asymmetries,
$A^\ell_{\rm FB}$ and $A^h_{\rm FB}$, are shown in Fig.~\ref{fig:ang} together with SM predictions~\cite{prediction}.
The statistical uncertainties are obtained using the likelihood-ratio ordering method~\cite{FC1}
and nuisance parameters are accounted for using the plug-in method~\cite{FC2}.
One dimensional 68\% Confidence Level (CL) intervals are obtained by varying one parameter at a time and treating
the others as nuisance parameters. In the analysis~\cite{lastLmumu} the statistical uncertainties on
$A^\ell_{\rm FB}$ and $f_{\rm L}$ are also reported in the form of two-dimensional 68\% CL regions,
where the likelihood-ratio ordering method is applied by varying both observables at the same time
and therefore taking correlations into account.

\section{Conclusions}

A measurement of the differential branching fraction of the rare $\Lambda_b^0 \rightarrow \Lambda \mu^+ \mu^-$ decay
is performed using data recorded by the LHCb detector at centre-of-mass energies of 7 and 8 TeV and
corresponding to an integrated luminosity of 3.0 fb$^{-1}$.
Evidence for the signal is found for the first time in the $q^2$ region
below the square of the $J/\psi$ mass, and in particular in the $0.1 < q^2 < 2.0$ GeV$^2/c^4$ interval,
where an enhanced yield is expected due to the proximity of the photon pole. The uncertainties of the measurements
in the $15 < q^2 < 20$ GeV$^2/c^4$ interval are reduced by a factor of approximately
three relative to the previous LHCb measurement~\cite{lhcbLmumu}. This improvement
is due to a larger data sample and a better control of systematic uncertainties.
The branching fraction measurements are compatible with SM predictions in the high-$q^2$ region,
above the square of the $J/\psi$ mass, and lie below the predictions in the low-$q^2$ region.
Furthermore, the first measurement of angular observables for this decay is reported.
Two forward-backward asymmetries, in the dimuon and $p\pi^{-}$ systems, are measured.
The measurements of the $A^h_{\rm FB}$ observable are in good agreement with the SM predictions while
for the $A^\ell_{\rm FB}$ observable measurements are consistently above the predictions.

\end{document}